# Non-Hermitian eigenvalue knots and their isotopic equivalence

Zhen Li[1], Kun Ding[2*], Guancong Ma[1*]

[1] Department of Physics, Hong Kong Baptist University, Kowloon Tong, Hong Kong, China

[2] Department of Physics, State Key Laboratory of Surface Physics, and Key Laboratory of Micro and Nano Photonic Structures (Ministry of Education), Fudan University, Shanghai 200438, China

*e-mail: kunding@fudan.edu.cn; phgcma@hkbu.edu.hk

**Abstract**

The spectrum of a non-Hermitian system generically forms a two-dimensional complex Riemannian manifold with distinct topology from the underlying parameter space. Spectral topology permits parametric loops to map the affiliated eigenvalue trajectories into knots. In this work, through analyzing exceptional points and their topology, we uncover the necessary considerations for constructing eigenvalue knots and establish their relation to spectral topology. Using an acoustic system with two periodic synthetic dimensions, we experimentally realize several knots with braid index 3. In addition, by highlighting the role of branch cuts on the eigenvalue manifolds, we show that eigenvalue knots produced by homotopic parametric loops are isotopic such that they can deform into one another by type-II or type-III Reidemeister moves. Our results not only provide a general recipe for constructing eigenvalue knots but also expand the current understanding of eigenvalue knots by showing that they contain information beyond that of the spectral topology.

**Introduction.** Since its incipience in the late 1990s, non-Hermitian formalism has proven to be a powerful method for analyzing open systems (1, 2). Non-Hermitian systems have been successfully realized across many realms, including optics, acoustics, mechanics, and quantum systems (3, 4). Perhaps the most intriguing and profound aspect of non-Hermitian systems is the generic existence of a complex-valued spectrum. This characteristic leads to an entirely new layer, spectral topology (5-8), which underpins many physical effects and phenomena unique to non-Hermitian systems. Examples include the emergence of defective spectral singularities known as exceptional points (EPs) (9-15), fractionally quantized Berry phases (16-18), non-Abelian state permutations (19), bulk Fermi arcs (20), non-Hermitian skin effects (21), and so on. Spectral topology can be diagnosed using topological invariants for eigenvalues (22, 23), which detects non-vanishing vorticity carried by spectral singularities. The non-trivial spectral topology permits the eigenvalues of different states to connect by crossing branch cuts, at which multiple sheets of the eigenvalue manifold intersect. Thus the possibility emerges for the states to permute, which can be captured by a permutation group (19, 24). During the permutation process, multiple eigenvalues braid into complex structures that become knots embedded in $\mathbb{R}^3$ (25-29). Recently, eigenvalue knots have been experimentally demonstrated in a two-state (28) and a three-state optical system (29).

The knot description is suitable for characterizing non-Hermitian bands because the fundamental group for an eigenvalue manifold is isomorphic to a braid group (5, 6). The braid algebra implies that the local perspective of a single EP is inadequate to characterize the collective behavior of multiple EPs, and a holistic view is essential. However, "knottiness" describes how closed loops are entangled when embedded in $\mathbb{R}^3$, and hence by definition, they must contain information beyond topology. For example, a trefoil knot and an unknot are topologically identical, or homeomorphic, because bijective maps between the two clearly exist. Conversely, different braid sequences can produce the same type of knot or isotopic knots, i.e., knots can transform into one another via Reidemeister moves (30-32). These peculiar properties of knots seem not entirely aligned with the description of spectral topology of non-Hermitian bands. Yet it remains unclear in what ways the knot description is different from spectral topology, and what advantage, if any, the eigenvalue knots description carries. As a result, the generic rules for constructing eigenvalue knots also remain obscure.

In this work, we present a systematic study and summarize the generic rules for constructing arbitrary eigenvalue knots with braid index $N = 3$. Considerations, including dimensional requirements of the parameter space, involvement of EPs and spectral branch cuts, and determination of parametric paths, are established based on their correspondences in knot theory. We then show how these generic rules guide the design and realization with two examples in a periodic parameter space: a figure-8 knot and a whitehead link using acoustic systems. Special attention is given to the role of spectral topology, and it is found that the eigenvalue winding numbers of EPs, which associate with the writhe number of the braid operations, are insufficient for the realization of specific eigenvalue knots. This finding aligns with the abovementioned fact that knots contain information beyond the attention of topology. Additionally, we highlight the relations among isotopic knots, Reidemeister moves, and homotopic loops in the parameter space. Our work clarifies the important role of knot theory for non-Hermitian spectral topology and builds the foundation for future investigations in this direction.

**General Considerations for Eigenvalue Knots.** We first lay out the general considerations for producing knots from complex eigenvalues. A knot is a series of crossings of strands that form one or multiple closed loops. (In some literature, the latter are called "links." But here, we follow the general knot theory and study all these cases together as knots.) It can be assigned with a braid index $N$ and a braid word (33). The braid index describes the least number of strands needed to form its braid diagram (34). The braid word is a sequence of braid operators $\tau_n^w$. The subscript $n$ is an integer that indicates the crossing between $n^{\text{th}}$ and $(n+1)^{\text{th}}$ strands, with $n \in [1, M-1]$ where $M$ is the total number of strands and $M \geq N$. (Note that it is legal for the number of braided strands $M$ to be larger than the braid index $N$ of the knot, and this has important consequences associated with type-I Reidemeister moves, as will be discussed later.) For $M = 3$, there are only two generating braid operators $\tau_1^1$ and $\tau_2^1$. The former (latter) occurs when the branch cut connecting states 1 and 2 (2 and 3) is crossed. The states are ordered ascendingly by real parts of their eigenvalues. It is straightforward to see that when an EP is encircled, the eigenvalue trajectory must cross at least one branch cut. The fact that $\tau_1^1$ and $\tau_2^1$ only exchange two neighbouring strands also implies the necessity to distinguish the states that form an order-2 EP (EP2). Hereafter, we denote the EP2 produced by the coalescence of states 1 and 2 (2 and 3) as $EP2_1$ ($EP2_2$), respectively. The superscript $w$ is the local writhe number. $w = +1(-1)$ represents that the $n^{\text{th}}$ strand over- (under-) crosses the $(n+1)^{\text{th}}$ strand. For simplicity, the superscript is omitted when $w = 1$. This implies a relation between the eigenvalue vorticity and the writhe number, which will be discussed later. The braid word is eventually determined by the sequence in which branch cuts are crossed.

Eigenvalue knots of $N = 3$ emerge only in systems with three or more bands. Since the generating braid operators only exchange neighbouring strands, we only need to focus on EP2s.

A three-state non-Hermitian Hamiltonian generically reads

$$H_{3b}(\boldsymbol{k}) = g_0(\boldsymbol{k})\lambda_0 + \boldsymbol{g}_\mu(\boldsymbol{k})\lambda_\mu, \tag{1}$$

where $\lambda_0$ is a 3 × 3 identity matrix and $\lambda_\mu$ are the Gell-Mann matrices with $\mu = 1,2,...8$, i.e., the extension of the Pauli matrices for SU(3). $g_0$ and $\boldsymbol{g}_\mu$ are complex functions of system parameter(s) $\boldsymbol{k}$. Our first consideration is that the conditions for EP2s to appear are $\text{Re}[\Delta] = 0$ and $\text{Im}[\Delta] = 0$, where $\Delta$ is the discriminant of the characteristic polynomial of Eq. (1) (*SI Appendix*). The vanishing of $\Delta$ imposes two independent constraints in the $\boldsymbol{k}$-space, each determining a $[\dim(\boldsymbol{k}) - 1]$-dimensional sub-manifold, and their intersections are the EPs. Thus, without imposing additional symmetry, the minimal dimension requirement for the stable existence of EP2s is $\dim(\boldsymbol{k}) = 2$ (35, 36). Furthermore, to stably form one-dimensional (1D) intersections, i.e., EP2 curves, the $\boldsymbol{k}$-space must be at least three-dimensional (3D).

The next consideration is the writhe number, where spectral topology plays a direct role. Each braid operation exchanges two neighboring bands, which can only be achieved by crossing a branch cut. And because any branch cut always connects two EP2s, the crossing direction is linked to the local winding around the EP. In other words, the topology of EP-hosting eigenvalue manifolds permits the braid operations, and such spectral topology can be captured by a local invariant called eigenvalue vorticity, which is defined for any pair of states $j$ and $j'$ as $v_{jj'}(\Gamma) = -\frac{1}{2\pi}\oint_\Gamma \nabla_{\boldsymbol{k}} \arg[\omega_j(\boldsymbol{k}) - \omega_{j'}(\boldsymbol{k})] \cdot d\boldsymbol{k}$, where $\Gamma$ denotes a parametric loop (22, 23). $v_{jj'}$ is nonzero only when $\Gamma$ encloses one or more Eps (22, 23, 37). Summing the vorticities $v_{jj'}$ for all combinations of states yields a net vorticity invariant (38) $\mathcal{V} = \sum_{j\neq j'} v_{jj'}(\Gamma)$. $\mathcal{V}$ directly corresponds to the total writhe number of an oriented braid (27) $\mathcal{V} = -W = -\sum w$, where the summation runs over all braid operators (31, 39). (To distinguish the writhe number of a knot from the local writhe number, we name the former as the total writhe number.)

Indeed, spectral topology has an important role in constructing an eigenvalue knot. Yet to obtain any specific knot, braid operations must be arranged in a specific sequence. We thus arrive at the last consideration: one must examine the order in which branch cuts are crossed to determine the order of the braid operators. This step is essential because, first, braid groups on $M > 2$ strands are non-Abelian. Second, the eigenvalue trajectory may cross a branch cut without encircling any EP, such crossings generate one of the Reidemeister moves, which change the braid words.

**Design and Realization of Eigenvalue Knots in a Periodic Non-Hermitian Model.** We are now well equipped to construct eigenvalue knots with $N = 3$. To demonstrate, consider the following three-state model

$$H(\phi_x, \phi_y, z) = (\omega_0 - i\gamma_0)\lambda_0 + \begin{bmatrix} (i+z)\alpha & -\kappa & 0 \\ -\kappa & F(\phi_x, \phi_y) & -\kappa \\ 0 & -\kappa & -(i+z)\alpha \end{bmatrix}, \quad (2)$$

where $\omega_0 - i\gamma_0$ is the onsite energy ($\omega_0 = 12905.4$ rad·s$^{-1}$, $\gamma_0 = 116.8$ rad·s$^{-1}$) and $\kappa = 40.8$ rad·s$^{-1}$ is the hopping strength. Herein, $i\alpha$ with $\alpha \in \mathbb{R}$ and $\text{Im}[F(\phi_x, \phi_y)]$, are non-Hermitian terms, and $F(\phi_x, \phi_y) = (a\cos^2\phi_x - ib\cos^2\phi_y) + C$ with $a, b \in \mathbb{R}$ and $C \in \mathbb{C}$. The parameters are $a = -291.9$ rad·s$^{-1}$, $b = 23.1$ rad·s$^{-1}$, and $C = 145.9 + 11.5i$ rad·s$^{-1}$. The values of these parameters are retrieved from experimental data using the Green's function (11). We first fix $\alpha = 69.4$ rad·s$^{-1}$ and $z = 0.8$. The form of $F(\phi_x, \phi_y)$ indicates that $\phi_{x,y} \in [0, 2\pi)$, i.e., the $\phi_x\phi_y$-subspace is homeomorphic to a 2-torus, as shown in Fig. 1*A*. We can identify sixteen EP2$_1$s and sixteen EP2$_2$s in the eigenvalue manifold (*SI Appendix*). Their positions and associated branch cuts on the $\phi_x\phi_y$-space are depicted in Fig. 1*A* (also in Fig. 3*A*). The sign of $\mathcal{V}$ for each EP2 is indicated nearby each EP2. The phase gradient of the discriminant, $\nabla_\phi[\arg(\Delta)]$, clearly shows the vorticity associated with each EP2 (Fig. 1*B*), is consistent with $\mathcal{V}$.

The two generating braid operators, $\tau_1$ and $\tau_2$, can be attained by counter-clockwise encircling one EP2$_1$ and one EP2$_2$ with $\mathcal{V} = -1$, respectively. (The inverse braid operators, $\tau_1^{-1}$ and $\tau_2^{-1}$, can be obtained by encircling the corresponding EP2 with an opposite $\mathcal{V}$ or by reversing the encircling direction.) Such loops are easily obtained by using the periodicity in the $\phi_x\phi_y$-space. In Fig. 1*A*, paths $\Gamma_1$ and $\Gamma_2$ are two possible choices. They each drive the eigenvalues to trace out trajectories on their manifold, which are unwrapped and shown by the solid lines in Fig. 1*C* and *D*. We choose the perspective that regards the left-hand side of the path as the interior. Path $\Gamma_1$ thus counter-clockwise encircles thirteen EP2$_1$s and twelve EP2$_2$s.

Of the thirteen EP2$_1$s, six have $\mathcal{V} = 1$ and seven have $-1$, so the net result is $\mathcal{V} = -1$. Moreover, the twelve EP2$_2$s annihilate pairwise. The net outcome is that the path $\Gamma_1$ equivalently encloses one EP2$_1$ with $\mathcal{V} = -1$ and no EP2$_2$, such that the first and second eigenvalues exchange, inducing the braid operator $\tau_1$ (Fig. 1*C*). By examining the branch cuts, it is straightforward to identify that $\Gamma_1$ transversely intersects with the branch cut of EP2$_1$s once. Following the same procedure, we can also identify that $\Gamma_2$ induces $\tau_2$ (Fig. 1*A* and *D*).

**Experimental Realization.** Next, we present an acoustic system to realize Eq. (2). Different from the acoustic cavity reported in previous works, our system here leverages the azimuthal degree of freedom in cylindrical cavities to realize periodic synthetic dimensions that effectively mimics the Bloch wavenumbers in lattice systems. Fig. 2*A* shows three identical air-filled cylindrical acoustic cavities stacked together (*Materials and Methods*). We focus on the second-order mode, whose pressure and velocity distributions are shown in Fig. 2*B*. The sound field is constant along the $z$-axis and varies along the azimuthal direction for a fixed radius, offering a natural realization of the two coordinates $\phi_x$ and $\phi_y$. (Fig. 2*A*. Therein, both the metal block and the sponge are placed in the top cavity for an easy photo, they are placed in the middle cavity in the experiments.) The radial nodal line perpendicular to the metal plate (Fig. 2*B*) is defined as $\phi_{x,y} = 0$. The offset to the real (imaginary) part of eigenfrequency is realized by the metal block (sponge) and is experimentally benchmarked, as shown in Fig. 2*C*. Both clearly follow a squared cosine function. We have also verified the offset effects in finite-element simulations (*Materials and Method*).

The experiments that demonstrate the braid operators $\tau_1$ and $\tau_2$ are performed in a stroboscopic manner. Specifically, we change the angular positions of the metal block and the sponge according to the parameters specified by paths $\Gamma_1$ and $\Gamma_2$, and measure the response spectra (both amplitudes and phases) in all three cavities at each synthetic coordinate. We then use the Green's function to retrieve the eigenfrequencies of the three modes (*SI Appendix*). The results are plotted by the dots in Fig. 1*C* and *D*, which clearly realize the intended operations.

With the generating braid operations established, we next construct eigenvalue knots of braid index 3. The large number of EP2s in conjunction with the 2-torus geometry of the parameter space offers plenty of choices. Two examples are shown in Fig. 3. We devise two different closed paths, denoted $\Gamma_3$ and $\Gamma_4$ (Fig. 3*A*). Path $\Gamma_3$ encloses fourteen EP2$_2$s and ten

EP2$_1$s, but only two EP2$_2$s with $\mathcal{V} = +1$ and two EP2$_1$s with $\mathcal{V} = -1$ remain after pairwise annihilation, such that the total writhe number is $W = -\mathcal{V} = 0$. By examining the sequence of branch cuts crossed by $\Gamma_3$, we conclude that the braid word produced is $\tau_2^{-1}\tau_1\tau_2^{-1}\tau_1$, which is a figure-8 knot (Fig. 3$B$). Likewise, $\Gamma_4$ produces the braid word $\tau_1\tau_2^{-1}\tau_1\tau_2^{-1}\tau_2^{-1}$, which has a common name of whitehead link (Fig. 3$C$). These two cases are experimentally realized using our acoustic-cavity system by following the designated paths, as shown in Fig. 3$B$ and $C$.

These results highlight an advantage of the knot description when multiple EP2s are involved. The figure-8 knot executes $\tau_2^{-1}\tau_1$ twice, while the whitehead link executes $\tau_1\tau_2^{-1}\tau_1$ and $\tau_2^{-1}\tau_2^{-1}$ in order. The $\tau_2^{-1}\tau_1$ and $\tau_1\tau_2^{-1}\tau_1$ are two concatenated operations that exchange all three eigenvalues, while $\tau_2^{-1}\tau_2^{-1}$ can be produced by two cycles around an EP2$_2$ with $|\mathcal{V}| = 1$ or one cycle around an EP2$_2$ with $|\mathcal{V}| = 2$ (19, 37). Therefore, the loop that produces the figure-8 knot is equivalent to a loop that encircles two order-3 EPs (EP3s) because an EP3 is the merger of at least one EP2$_1$ and one EP2$_2$. Likewise, the whitehead link can equivalently be produced by a loop that sequentially encloses an EP3 and the EP2$_2$s with a net $|\mathcal{V}| = 2$. In other words, knots produced by homotopic loops are isotopic, meaning they are essentially the same (31). However, this observation brings us to another question.

From the topological perspective, eigenvalue details around a continuous exceptional curve does not matter. However, this clearly is not true for knots because homotopic eigenvalue trajectories may cross branch cuts differently, thus the corresponding knots must have different braid words. What are the implications of this change in braid words?

**Isotopic Knots, Homotopic Paths, and Reidemeister Moves.** To answer, we re-examine the parameter $z$ in Eq. (2), which extrudes the 2-torus spanned by $\phi_x$ and $\phi_y$ into the third dimension. As mentioned, EP2s can stably form 1D curves in a 3D space. These EP2 curves are shown in Fig. 4$A$. Because $z$ tunes the onsite energy of sites 1 and 3 and its presence respects inversion symmetry, the states forming the EP2s are flipped in their real energy when $z$ changes the sign. Because all the EP2 curves remain continuous and smooth across $z = 0$, their vorticity invariants remain the same (*SI Appendix*), the change in the type of the EP2s cannot be straightforwardly diagnosed unless using the eigenvalue knots. We illustrate this via the two homotopic loops, denoted paths $\Gamma_5$ and $\Gamma_6$ in Fig. 4$A$. Path $\Gamma_5$ produces $\tau_2^{-1}\tau_1\tau_1\tau_2$, which is a Hopf link formed by the first and the third states and an unknot from the second state (Fig. 4*B*). Path $\Gamma_6$ is obtained by translating $\Gamma_5$ from $z = 0.2$ to $-0.2$ without crossing any EP2s. However, because all EP2$_1$s become EP2$_2$s and vice versa, the eigenvalue knot has a different braid word $\tau_2\tau_1^{-1}\tau_1\tau_2\tau_1^{-1}\tau_1$, which is simplified to $\tau_2\tau_2$ by two type-II Reidemeister moves (Fig. 4*C*). The result is a Hopf link and an unknot. In other words, the knots produced by $\Gamma_5$ and $\Gamma_6$ are isotopic, which reveals an intriguing connection between homotopic loops, isotopic eigenvalue knots, and Reidemeister moves (31, 32).

Reidemeister moves consist of three local manoeuvres through which one knot can be transformed into its isotopic form. There are three types of Reidemeister moves: type-I "twist," type-II "poke," and type-III "slide." One would expect the homotopic transformation of parametric loops can produce all three types of moves. However, our results indicate otherwise.

We first illustrate the two possible moves, type-II and type-III. Type-II move adds to a braid word two consecutive braid operations that cancel each other, i.e., it adds an identity $\tau_n\tau_n^{-1} = e$ to a braid word (40). To demonstrate its appearance, we design two homotopic loops in the $\phi_x\phi_y z$-space, as shown in Fig. 5*A*. Loop $\Gamma_C$ encircles no EP2 and does not cross any branch cut, and the three bands produce an unlink (Fig. 5*C(i)*). Loop $\Gamma_D$ is homotopic to $\Gamma_C$ but crosses the branch cut of EP2$_2$ twice. Hence the braid word is $\tau_2\tau_2^{-1}$ (Fig. 5*C(ii)*). In other words, a type-II move is generated by two consecutive crossings of the same branch cut.

Type-III move formally appears as $\tau_n\tau_{n+1}^{\pm 1}\tau_n^{-1} = \tau_{n+1}^{-1}\tau_n^{\pm 1}\tau_{n+1}$, which corresponds to moving a stand-alone strand across one braid (32). This is attained by translating a loop such that the encircled EP2 changes types. In Fig. 5*A*, loops $\Gamma_E$ and $\Gamma_F$, belonging to the same equivalence class, fulfill these requirements. Loop $\Gamma_E$ counter-clockwise encircles one EP2$_1$ with $\mathcal{V} = -1$, and it produces $\tau_2^{-1}\tau_1\tau_2$, as shown in Fig. 5*D(i)*. Loop $\Gamma_F$ is obtained by translating $\Gamma_E$ to $z = -0.2$, the generated braid word is $\tau_1\tau_2\tau_1^{-1}$ (Fig. 5*D(ii)*), i.e., the two braid words transform by a type-III move.

Surprisingly, the type-I move, which adds or removes a local twist, cannot appear for two homotopic loops. The reason is transparent: type-I move does not preserve the total writhe number. In our context, it means the net vorticity of the eigenvalue trajectory must change, which can only occur when the number of the encircled EP2s is different. Loops $\Gamma_A$ and $\Gamma_B$ in Fig. 5A are an example. $\Gamma_A$ encloses no EP2, and each band forms an unknot ($N = 1$), such that the writhe numbers are zero and no twisting is seen (Fig. 5B(*i*)). $\Gamma_B$ encloses one EP2$_1$, and the first and second bands together form an unknot, i.e., we have $M = 2$ but $N = 1$. This unknot has a twist, and the total writhe number is one, as shown in Fig. 5B(*ii*). In fact, such a twist generated by type-I move naturally appears when the braid index $N$ is smaller than the number of the involved bands $M$. The reason is straightforward: the only option for two bands to join in the eigenvalue manifold is for the eigenvalue trajectory to cross a branch cut, which also corresponds to a single braid operation, thus producing a twist.

**Discussion**

Our study has clarified several important issues about non-Hermitian eigenvalue knots. While spectral topology of the eigenvalue manifold is indeed the fundamental reason that eigenvalues can braid into knots, the intrinsic richness of knot structures means they must contain information beyond topology, which is largely ignored in previous studies. We have shown that such richness manifests in two aspects. First, specific braid words can only be determined by the full picture of the eigenvalue manifold, including branch cuts and EP types, which are obviously not "topological knowledge" of the manifold. By fully accounting for all these aspects, we have arrived at a set of generic considerations that are useful in designing specific knot structures. Their effectiveness is demonstrated in our experimental realizations of complex knots previously beyond experimental reach. Second, eigenvalue trajectories that are considered topologically equivalent, or homotopic, can in fact correspond to different braid words describing isotopic knots, which are connected via Reidemeister moves. To conclude, our results and findings constitute a foundation for the future study of non-Hermitian spectral behaviors, which are essential in understanding novel phenomena such as non-Hermitian skin effect (27), non-Hermitian state permutations (19), and non-Hermitian topological phase transition (25).

**Materials and Methods**

**Realization of Synthetic Dimension.** The acoustic system is composed of three identical cavities with a radius of 5 cm and height of 4 cm (Fig. 2*A*). A thin metal plate is inserted to break the circular symmetry by reflecting the clockwise and counter-clockwise circulating waves in the cavity, such that stable standing-wave eigenmodes are formed. The sound field in our cylindrical cavity only varies in the radial and azimuthal directions and is constant along the $z$-direction. For the second-order mode that we treat as the onsite orbital, at a fixed radius, the pressure follows $P(\phi) \propto \cos\phi$. This characteristic is leveraged for realizing the synthetic dimensions. The first synthetic coordinate $\phi_x$ tunes the real part of the onsite eigenfrequency. This is achieved by placing a small block of metal on the inner wall of the cavity. The metal block's perturbation to the real eigenfrequency is sensitive to the local intensity of the pressure, thus it follows $\cos^2\phi_x$. This is confirmed by benchmarking the real detuning as a function of $\phi_x$. The real detuning is retrieved using the Green's function (*SI Appendix*), and the results are shown in Fig. 2C. Thus, the azimuthal position of the metal block plays the role of $\phi_x$. It is noticed that the real part of the eigenfrequency attains the peak value when the metal block is placed at $\phi_x = \frac{\pi}{2}/\frac{3\pi}{2}$. This is because the potential energy is increased with the compressible fluid region decreasing, whereas the kinetic energy is hardly affected. Therefore, the eigenfrequency is enlarged according to Rayleigh's method (41-43). When the metal block is placed at the nodal line of the second order mode, the velocity is strengthened locally such that the kinetic energy increases, while the potential energy remains unchanged, therefore, the eigenfrequency decreases to the minimum values. The distribution acoustic velocity field $v$ is leveraged to tune the dissipation, which realizes the second synthetic dimension $\phi_y$. Because acoustic velocity is related to pressure via $\rho_0 \frac{\partial v}{\partial t} = -\nabla P$ with $\rho_0$ being the density of air, the maximal (minimal) velocity appears at the position where the pressure is zero (maximal) (Fig. 2B). The dissipation induced by an anechoic sponge is proportional to local kinetic energy, thus quadratic in acoustic velocity. Hence the azimuthal position of the sponge tunes the loss in the cavity, as shown in Fig. 2C. Thus, the azimuthal position of the sponge realizes $\phi_y$. The hopping terms in Eq. (2) are realized by coupling the neighboring cavities with an 18.9-mm$^2$ hole. In the experiment, the top cavity is excited by a loudspeaker connecting to the rim via a small opening. A microphone is inserted into each cavity such that it is flush with the inner rims to measure the sound pressure.

**Finite-element Method.** To further verify our scheme, we have also performed numerical simulations using the commercial finite-element solver COMSOL Multiphysics (v5.4). The metal block is modeled as protrusion into the cavity at the boundary, and the sponge is modeled as a poroacoustic region using the Delany-Bazley-Miki model, with flow resistivity of $R_f = 1.8 \times 10^4 \text{ Pa} \cdot \text{s/m}^2$ (the value is obtained by fitting the FWHM of the experimentally measured response). Excellent agreement with experimental results is obtained (Fig. 2*C*).

**ACKNOWLEDGMENTS.** This work is supported by the National Natural Science Foundation of China (11922416, 12174072), the Hong Kong Research Grants Council (RFS2223-2S01, 12302420, 12300419, 12301822), and the Natural Science Foundation of Shanghai (No. 21ZR1403700).

**Author contributions:** Z. L. performed numerical and experimental studies. All authors engaged in the theoretical studies, the analysis of data, and the writing of the manuscript. G. M. conceived the research.

**Competing interests Statement:** The authors declare no competing interests.

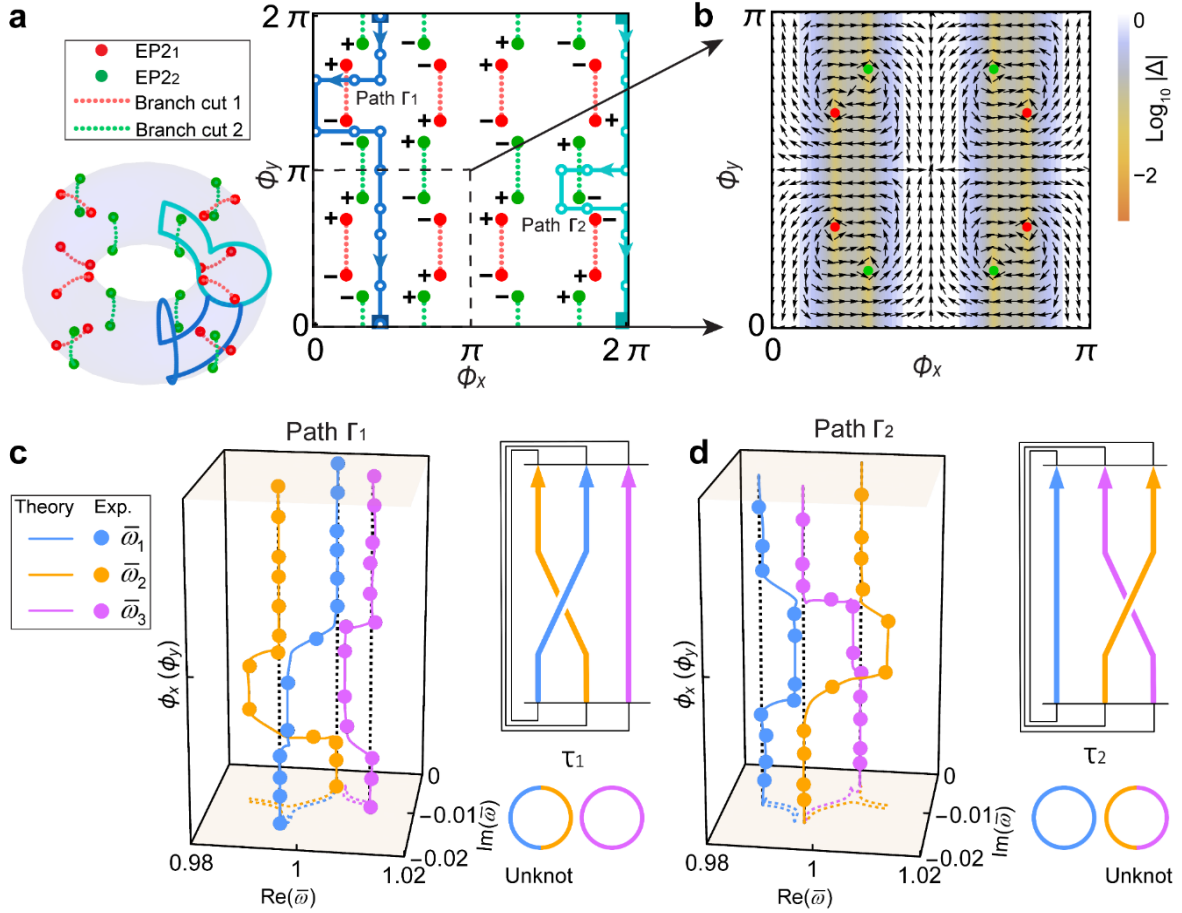

**Fig. 1.** EP2 features and braid operators. (*A*) EP2s and branch cuts projected on the $\phi_x\phi_y$-space at $z = 0.8$, which is equivalent to a 2-torus (left). EP2$_1$s (EP2$_2$s) are depicted by the red (green) dots, with the positive (negative) signs indicating $\mathcal{V} = 1\ (-1)$. The dotted red and green lines are the branch cuts that connect the two EP2s. The solid blue and cyan lines represent two closed paths, $\Gamma_1$ and $\Gamma_2$. (*B*) The color map and arrows show the discriminant norm in a log scale and its phase gradient within the dashed black box in *A*, in which the eigenvalue vorticities are clearly seen. (*C* and *D*) The evolutions of eigenvalues $\bar{\omega} = \omega_j/\omega_0\ (j = 1, 2, 3)$ along $\Gamma_1$ and $\Gamma_2$ trace out the two generating braid operations $C\ \tau_1$ and $D\ \tau_2$, wherein the curves (dots) are theoretical (experimental) results. The right panels of plot the braid diagram and the corresponding knots. The experimental parametric positions are denoted by the open circles in *A*, and the squares denote the start and end points.

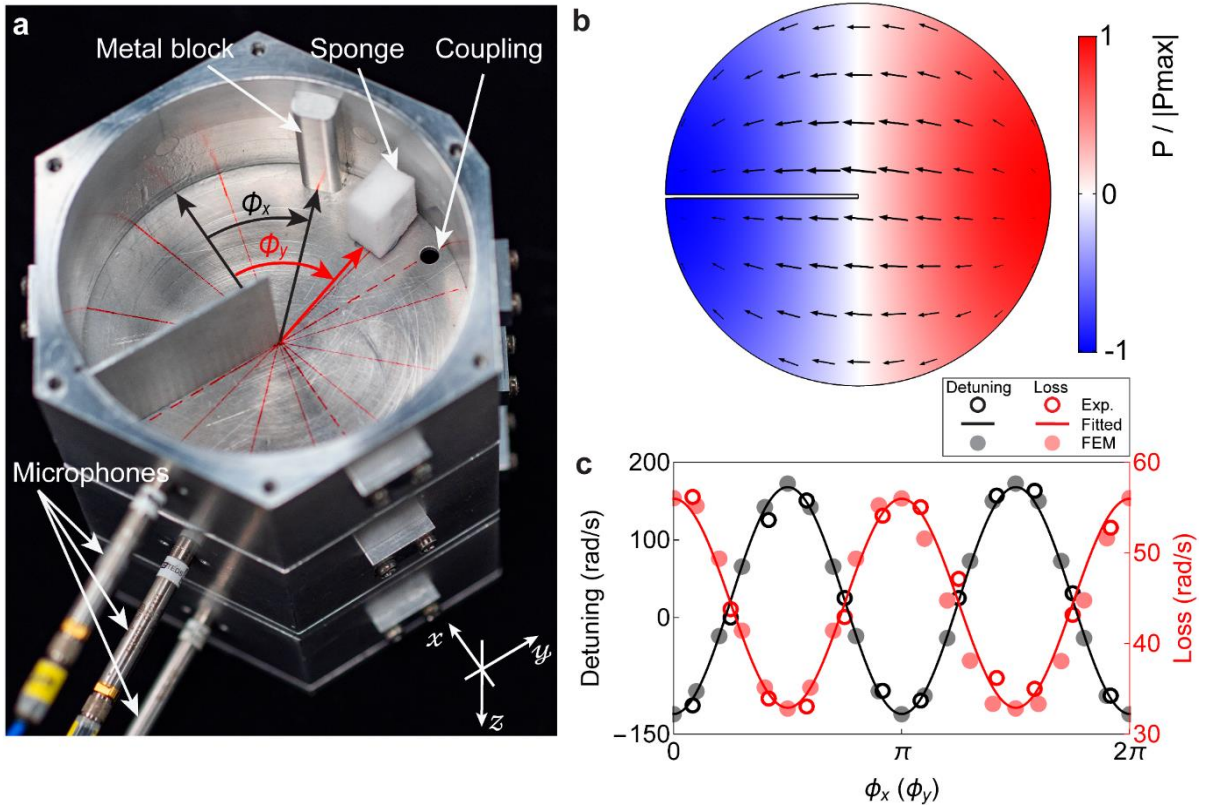

**Fig. 2.** Acoustic-cavity system. (*A*) The photo of the acoustic cavities. The metal block (acoustic sponge) with its angular position directly realizing $\phi_x$ ($\phi_y$) achieves the onsite detuning (loss). The hole at $\pi/2$ provides the coupling between cavities, and the microphones are used for measurements. (*B*) The distribution of acoustic pressure (color map) and velocity field (arrows) of the second-order mode for a single cavity. The arrow length is proportional to the magnitude of velocity. (*C*) The measured (simulated) detuning and loss as functions of $\phi_x$ and $\phi_y$ are shown by the empty (solid) black and red dots, respectively. The solid lines depict the results fitted by the function $F(\phi_x, \phi_y)$.

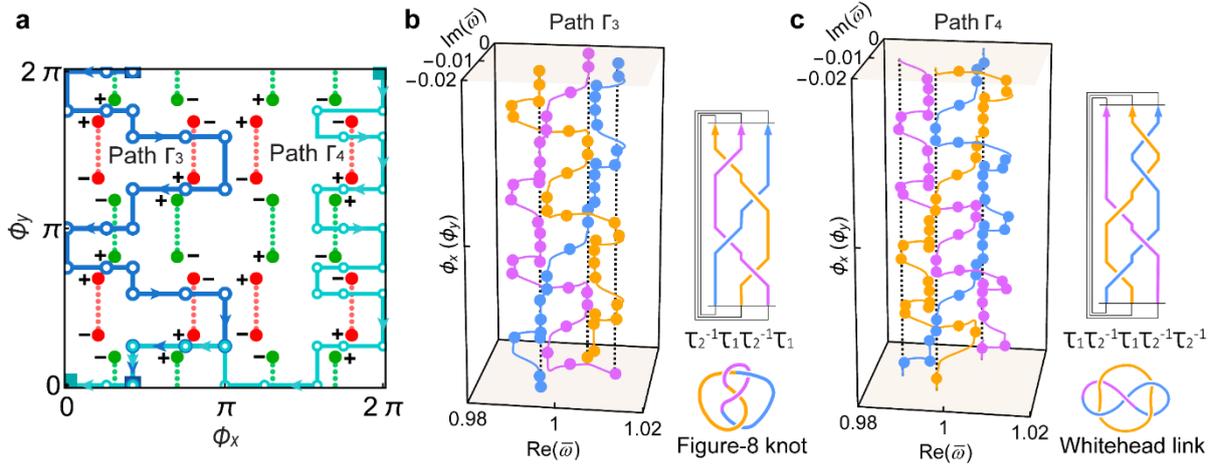

**Fig. 3.** Eigenvalue knots with braid index 3. (*A*) Two tailored closed paths $\Gamma_3$ and $\Gamma_4$ in the $\phi_x\phi_y$-space at $z = 0.8$. (*B* and *C*) The evolution of eigenvalues $\bar{\omega}$ along $\Gamma_3$ ($\Gamma_4$) generates a figure-8 knot (whitehead link). The curves and dots in the left panels of *B* and *C* are theoretical and experimental results, respectively. The right panels show the braid diagrams, the braid words, and the corresponding knots.

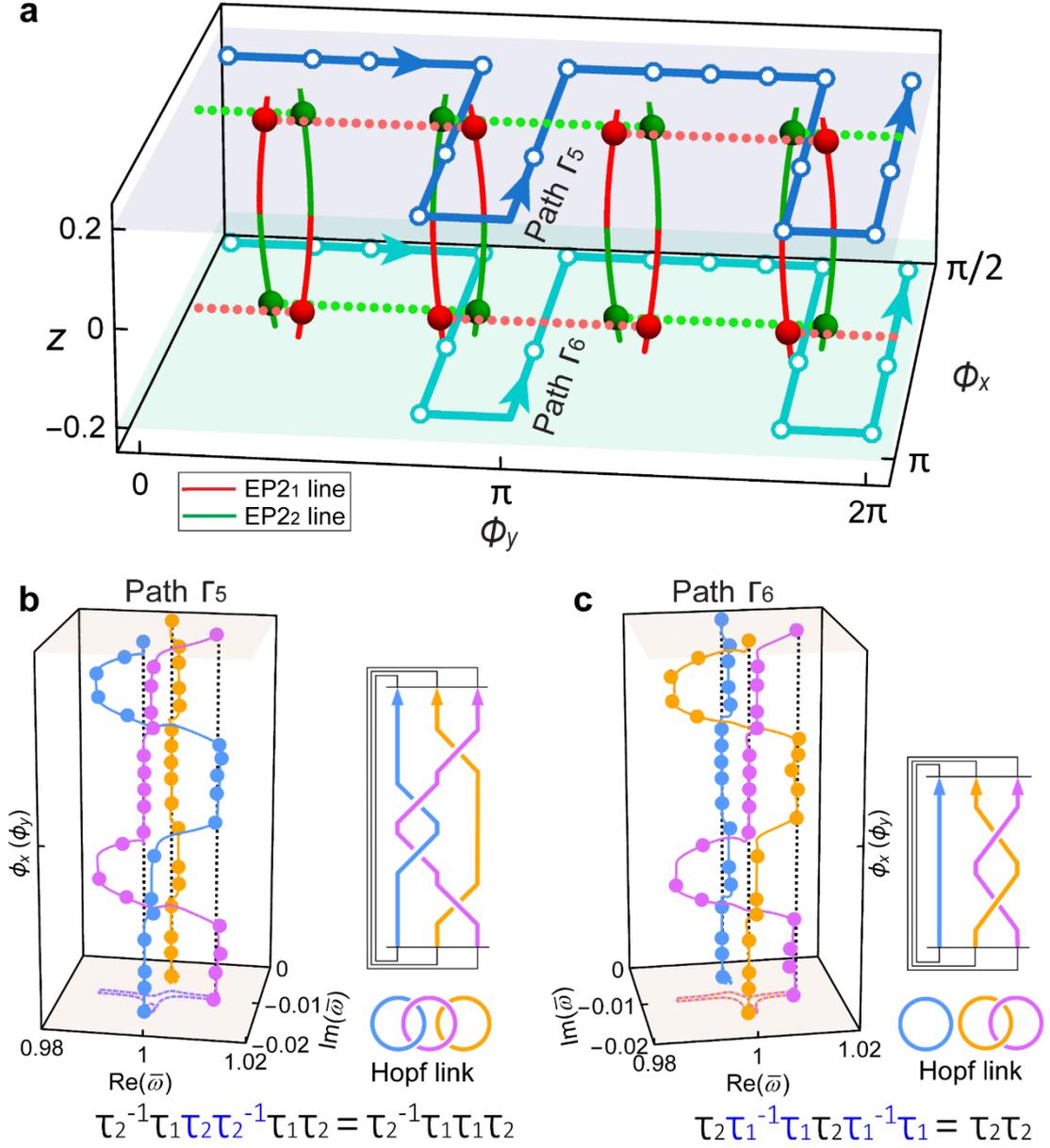

**Fig. 4.** Homotopic loops and isotopic knots. (*A*) The EP2s form continuous curves in the $\phi_x\phi_y z$-space, with red and green indicating EP2$_1$ and EP2$_2$, respectively. The dotted lines are branch cuts. Two homotopic loops, $\Gamma_5$ and $\Gamma_6$, are shown in blue and cyan. (*B* and *C*) The eigenvalue knots produced by $\Gamma_5$ and $\Gamma_6$ are experimentally (theoretically) illustrated in the left panel by the dots (curves), with the braiding diagrams and the corresponding knots sketched on the right. The braid words directly from the evolutions and the ensuing clean ones are shown at the bottom.

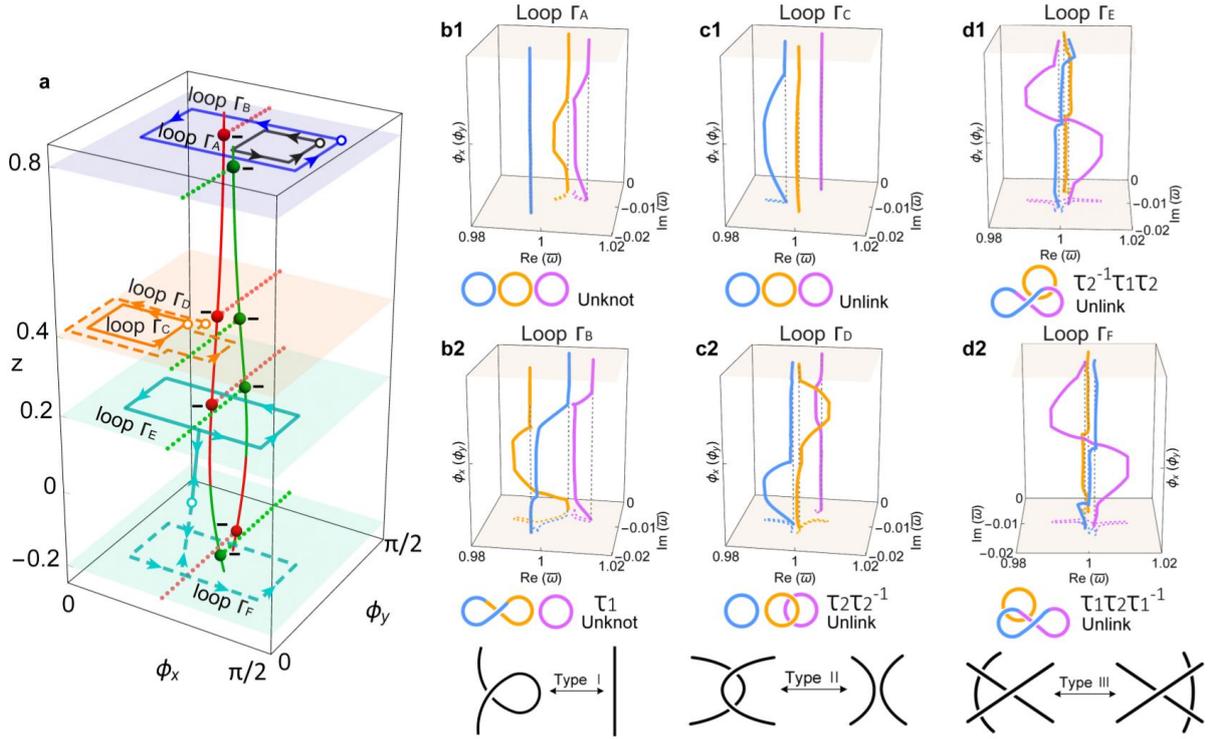

**Fig. 5.** Reidemeister moves appearing in eigenvalue braids. (*A*) The EP2s form continuous curves in the $\phi_x\phi_y z$-space, with red and green indicating EP2$_1$ and EP2$_2$, respectively. The dotted lines are branch cuts. Loops $\Gamma_A$ and $\Gamma_B$ are shown in black and blue on the $\phi_x\phi_y$-plane (blue in **a**) at $z = 0.8$. Loops $\Gamma_C$ and $\Gamma_D$ ($\Gamma_E$ and $\Gamma_F$) are homotopic, as shown in solid and dashed orange (cyan) on the $\phi_x\phi_y$-plane at $z = 0.4$ ($z = \pm 0.2$). (*B*) The eigenvalue knots formed by *B*(*i*) $\Gamma_A$ and *B*(*ii*) $\Gamma_B$ are related by a type-I move. (*C*) The eigenvalue knots produced by *C*(*i*) $\Gamma_C$ and *C*(*ii*) $\Gamma_D$ are unlinks related by a type-II move. (*D*) The eigenvalue knots induced by *D*(*i*) $\Gamma_E$ and *D*(*ii*) $\Gamma_F$ are unlinks of related by a type-III move.

# Supplementary Materials

## Non-Hermitian eigenvalue knots and their isotopic equivalence

Zhen Li[1], Kun Ding[2], Guancong Ma[1]

[1]Department of Physics, Hong Kong Baptist University, Kowloon Tong, Hong Kong, China

[2] Department of Physics, State Key Laboratory of Surface Physics, and Key Laboratory of Micro and Nano Photonic Structures (Ministry of Education), Fudan University, Shanghai 200438, China

**Contents**

## Section 1. Mathematics of the generic model

A three-state non-Hermitian Hamiltonian generically reads

$$H_{3b}(\mathbf{k}) = g_0(\mathbf{k})\lambda_0 + \boldsymbol{g}_\mu(\mathbf{k})\boldsymbol{\lambda}_\mu, \tag{S1}$$

where $\lambda_0$ is a $3 \times 3$ identity matrix, $\mu = (1, \ldots 8)$, and $\boldsymbol{\lambda}_\mu$ are the eight Gell-Mann matrices, i.e., the extension of the Pauli matrices for SU(3). $g_0$ and $\boldsymbol{g}_\mu$ are complex functions of system parameter(s) $\mathbf{k}$. The characteristic polynomial of the Hamiltonian reads

$$f_{\mathbf{k}}(\omega) = -\omega^3 + 3g_0\omega^2 + b(\mathbf{k})\omega + c(\mathbf{k}), \tag{S2}$$

where $b(\mathbf{k}) = -3g_0^2 + \sum_{\mu=1}^{8} g_\mu^2$ and $c(\mathbf{k}) = g_0^3 - g_0 \sum_{\mu=1}^{8} g_\mu^2 + g_3(g_4^2 + g_5^2 - g_6^2 - g_7^2) + 2(g_2 g_5 g_6 - g_2 g_4 g_7 + g_1 g_4 g_6 + g_1 g_5 g_7) + \frac{1}{3\sqrt{3}}[6g_8(g_1^2 + g_2^2 + g_3^2) - 3g_8(g_4^2 + g_5^2 + g_6^2 + g_7^2) - 2g_8^3]$. The discriminant of the polynomial is

$$\Delta_f(\omega) = 4b(\mathbf{k})^3 - 27c(\mathbf{k})^2 - 54b(\mathbf{k})c(\mathbf{k})g_0 + 9b(\mathbf{k})^2 g_0^2 - 108c(\mathbf{k})g_0^3 = 0. \tag{S3}$$

The conditions for the coalescence of two eigenstates are $\text{Re}[\Delta_f(\omega)] = 0$ and $\text{Im}[\Delta_f(\omega)] = 0$. We also remark that the stable existence of EP3s in the absence of additional symmetry requires $\dim(\boldsymbol{k}) \geq 4$. More generally, $2(n-1)$ real constraints need to be satisfied to find the EP$n$ in an $n$-state system[1].

## Section 2. The three-state model in the main text

Here, we present more details about the three-state model in the main text. The coefficients of the Gell-Mann matrices are parameterized by three synthetic dimensions $\{\phi_x, \phi_y, z\}$ as $g_0 = (\omega_0 - i\gamma_0) + F(\phi_x, \phi_y)/3$, $g_1 = g_6 = -\kappa$, $g_3 = [\alpha(i+z) - F(\phi_x, \phi_y)]/2$, $g_8 = \sqrt{3}[\alpha(i+z)/2 + F(\phi_x, \phi_y)/6]$, and $g_\mu = 0$ for other values of $\mu$, in which $F(\phi_x, \phi_y) = (a\cos^2\phi_x - ib\cos^2\phi_y) + C$ with $a, b \in \mathbb{R}$ and $C \in \mathbb{C}$. Therefore, the model reads

$$H(\phi_x, \phi_y, z) = (\omega_0 - i\gamma_0)\lambda_0 + \begin{bmatrix} (i+z)\alpha & -\kappa & 0 \\ -\kappa & F(\phi_x, \phi_y) & -\kappa \\ 0 & -\kappa & -(i+z)\alpha \end{bmatrix}, \tag{S4}$$

where $\omega_0 = 12905.4 \text{ rad} \cdot \text{s}^{-1}$, $\gamma_0 = 116.8 \text{ rad} \cdot \text{s}^{-1}$, $\kappa = 40.8 \text{ rad} \cdot \text{s}^{-1}$, and $\alpha = 69.4 \text{ rad} \cdot \text{s}^{-1}$. Therein, we fix $z = 0.8$ in Fig. 1 and Fig. 3, and $z = \pm 0.2$ in Fig. 4.

When $z = 0.8$, the positions of two types of EP2s formed by the intersection of the real and imaginary parts of the discriminant are shown in Fig. S1(a). Figure S1(b) shows the real parts of complex eigenvalues on the $\phi_x\phi_y$-plane, where the two types of EP2 can be clearly distinguished from the states they are produced.

Figure S2(a) shows the EP2 lines when $z$ is treated as the third synthetic dimension. The two different types of EP2s can be distinguished by examining the real parts of complex eigenvalues on the $\phi_x\phi_y$-plane at different $z$ [Fig. S2(b-d)]. We observe that the coalescing states of the EP2s change when $z$ varies from positive to negative values. At $z = 0$, the branch cuts stemming from the EP2s coincide.

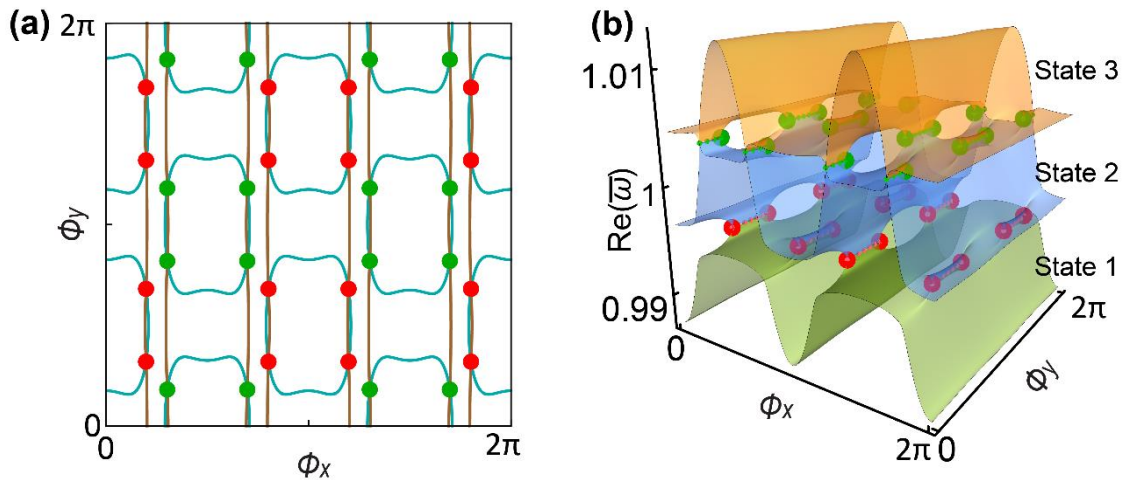

**Fig. S1.** Positions of the EP2s on the $\phi_x\phi_y$-plane at $z = 0.8$. (a) EP2s formed by the intersection of the real (cyan curves) and imaginary parts (brown curves) of the discriminant $\Delta_f = 0$. The red (green) points represent the EP2$_1$s (EP2$_2$s). (b) The real parts of the complex eigenvalues on the $\phi_x\phi_y$-plane. The states are ordered by $\text{Re}(\bar{\omega})$ from lower to higher frequency.

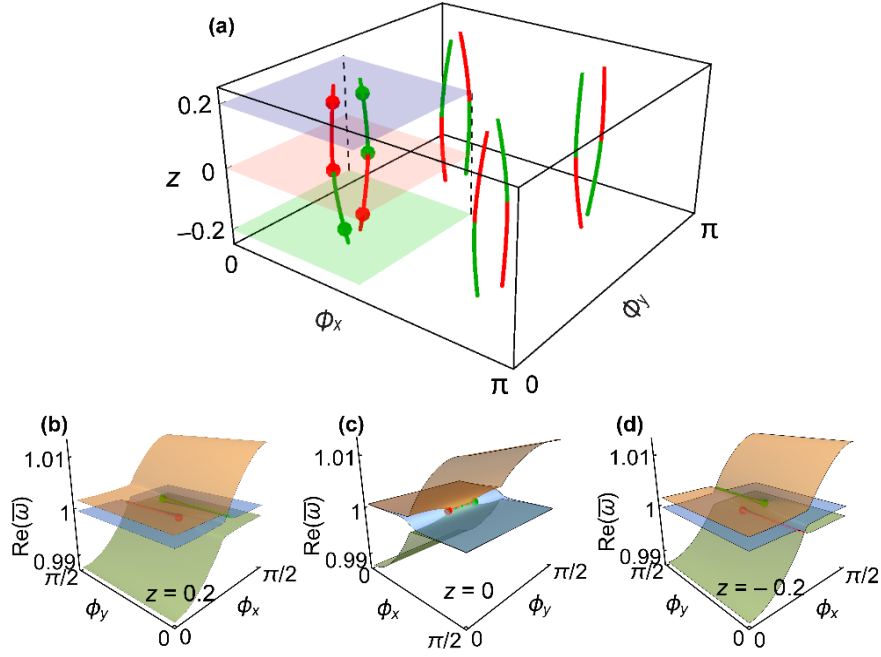

**Fig. S2.** (a) The EP2 curves in the $\phi_x\phi_y z$-space. (b), (c), (d) The real parts of the eigenvalues on the $\phi_x\phi_y$-plane at $z = 0.2, 0, -0.2$, respectively. The red (green) points represent EP2$_1$s (EP2$_2$s) formed by the coalescence of states 1 and 2 (states 2 and 3). The dotted red and green lines are the branch cuts.

Paths $\Gamma_5$ and $\Gamma_6$ in the main text are plotted on their respective $z$-plane in Fig. S3(a) and (c). In Fig. S3(a), we observe that path $\Gamma_5$ crosses branch cuts six times as it encircles two EP2$_1$s, which induces the braid word $\tau_2^{-1}\tau_1\tau_2\tau_2^{-1}\tau_1\tau_2$. Path $\Gamma_5$ becomes $\Gamma_6$ when it is translated from $z = 0.2$ to $z = -0.2$. In Fig. S3(c), we can identify that the braid word is changed to $\tau_2\tau_1^{-1}\tau_1\tau_2\tau_1^{-1}\tau_1 = \tau_2\tau_2$. On the other hand, it can be seen that vorticity invariants $\mathcal{V}$ for all the EP2s are not affected by the variation of $z$ [Fig. S3(b) and (d)]. These results show that homotopic paths can produce isotopic eigenvalue knots formed by different bands.

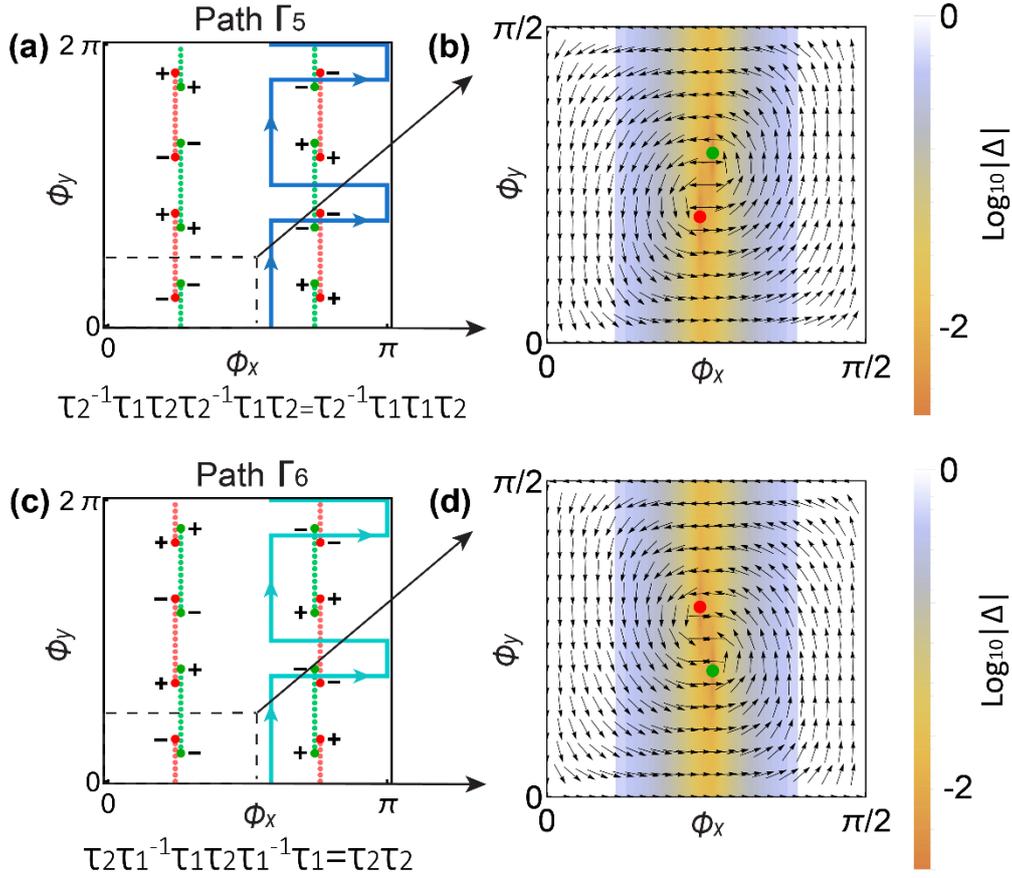

**Fig. S3.** (a, c) The positions of the $EP2_1$ (red) and $EP2_2$ (green) projected on the $\phi_x\phi_y$-plane (a) $\Gamma_5$ (solid blue line) on the $\phi_x\phi_y$-plane at $z = 0.2$ (c) $\Gamma_6$ (solid cyan line) on the $\phi_x\phi_y$-plane at $z = -0.2$. The positive (negative) signs indicate the $\mathcal{V} = 1$ ($\mathcal{V} = -1$). The dotted red and green lines are branch cuts. (b, d) The color maps and arrow fields show the discriminant norm in log scale and gradient $\nabla_\phi[\text{Im}(\ln \Delta)]$ for the region enclosed by the dashed black box in (a, c), respectively.

**Section 3. Retrieval of parameters**

The pressure response of an isolated cavity is first measured. The response is fitted using the function $P(\omega) = c/[\omega - (\omega_0 - i\gamma_0)]$, which is essentially the Green's function for a single mode, to obtain the resonant frequency $\omega_0$ and the intrinsic loss $\gamma_0$, where $c$ is a coefficient[2]. Then the detuning and loss effects are benchmarked in the single cavity before performing the subsequent experiment. Cases with the metal block (sponge) being placed at different azimuthal angles are tested and fitted, and the data is plotted in Fig. 2(c).

By fitting the 12 sets of data in Fig. 2(c), we obtain the detuning $X(\phi_x) = -291.9\cos^2\phi_x + 167.9 \text{ rad}\cdot\text{s}^{-1}$ and loss $Y(\phi_y) = 23.1\cos^2\phi_y + 32.9 \text{ rad}\cdot\text{s}^{-1}$. Based on the expressions of detuning and loss, we build the function $F(\phi_x, \phi_y) = (a\cos^2\phi_x - ib\cos^2\phi_y) + C = (X + x_0) - (Y + y_0)i$, where $x_0 = -22$ rad·s$^{-1}$ and $y_0 = -44.4$ rad·s$^{-1}$ are the initial offset on the detuning and loss, thus $C = 145.9 + 11.5i$ rad·s$^{-1}$.

For the experiments with the three-cavity system, the real and imaginary parts of the eigenfrequencies of the coupled modes can be obtained from the pressure responses using the Green's function

$$\vec{G}(\omega) = \sum_{j=1}^{3} \frac{|\tilde{\phi}_j^R\rangle\langle\tilde{\phi}_j^L|}{\omega - \omega_j} \quad (S5)$$

where $|\tilde{\phi}_j^R\rangle$ and $\langle\tilde{\phi}_j^L|$ are the normalized biorthogonal right and left eigenvectors, and $\omega_j$ are the eigenvalues of the Hamiltonian. In our experiments, because there is only one microphone in each cavity, $|\tilde{\phi}_j^R\rangle$ ($\langle\tilde{\phi}_j^L|$) is a three-entry column (row) complex vector. For more details of the retrieval methods, we refer interested readers to the Supplementary texts of refs.[3,4].

Tables S1-S5 show the offset on the real (imaginary) part of the onsite resonant frequency of cavity-2 when the positions of the metal block and sponge are changed along the closed path $\Gamma_1$ to path $\Gamma_6$, respectively. The points are numbered in the order they appear in the encircling paths. For each parametric path, two sets of data are chosen, and the experimentally measured spectrum, as well as the fitted results, are plotted in Fig. S4.

Table S1. Parameters on path $\Gamma_1$.

| Point No. | $\phi_x$ (rad) | $\phi_y$ (rad) | X (rad/s) | Y (rad/s) |
|---|---|---|---|---|
| 1 | 1.31 | 6.28 | 150.01 | 51.39 |
| 2 | 1.31 | 5.50 | 150.01 | 46.18 |
| 3 | 1.31 | 4.97 | 148.22 | 36.19 |
| 4 | 0.79 | 4.97 | 25.16 | 30.12 |
| 5 | 0.00 | 4.97 | -122.94 | 34.79 |
| 6 | 0.00 | 3.93 | -115.76 | 46.90 |
| 7 | 0.79 | 3.93 | 29.59 | 46.48 |
| 8 | 1.31 | 3.93 | 156.44 | 48.74 |
| 9 | 1.31 | 3.14 | 142.58 | 51.54 |
| 10 | 1.31 | 2.36 | 139.83 | 44.98 |
| 11 | 1.31 | 1.83 | 143.35 | 38.07 |
| 12 | 1.31 | 0.79 | 150.05 | 47.92 |
| 13 | 1.31 | 0.00 | 150.01 | 51.39 |

Table S2. Parameters on path $\Gamma_2$.

| Point No. | $\phi_x$ (rad) | $\phi_y$ (rad) | X (rad/s) | Y (rad/s) |
|---|---|---|---|---|
| 1 | 6.28 | 5.50 | -122.08 | 47.44 |
| 2 | 6.28 | 4.97 | -122.94 | 34.79 |
| 3 | 6.28 | 3.93 | -115.76 | 46.90 |
| 4 | 6.28 | 3.14 | -125.87 | 55.51 |
| 5 | 5.50 | 3.14 | 29.25 | 50.58 |
| 6 | 4.97 | 3.14 | 146.22 | 60.48 |
| 7 | 4.97 | 2.36 | 147.39 | 45.34 |
| 8 | 5.50 | 2.36 | 28.16 | 38.99 |
| 9 | 6.28 | 2.36 | -122.96 | 43.33 |
| 10 | 6.28 | 1.83 | -127.55 | 30.02 |
| 11 | 6.28 | 0.79 | -126.94 | 51.91 |

Table S3. Parameters on path $\Gamma_3$.

| Points No. | $\phi_x$ (rad) | $\phi_y$ (rad) | X (rad/s) | Y (rad/s) |
|---|---|---|---|---|
| 1 | 1.31 | 6.28 | 150.01 | 51.39 |
| 2 | 0.79 | 6.28 | 27.31 | 50.89 |
| 3 | 0.00 | 5.50 | -122.08 | 47.44 |
| 4 | 0.79 | 5.50 | 17.15 | 43.85 |
| 5 | 1.31 | 5.50 | 150.01 | 46.18 |
| 6 | 1.31 | 4.97 | 148.22 | 36.19 |
| 7 | 2.36 | 4.97 | 24.16 | 41.22 |
| 8 | 3.14 | 4.97 | -131.29 | 39.49 |
| 9 | 3.14 | 3.93 | -130.11 | 36.89 |
| 10 | 2.36 | 3.93 | 28.12 | 38.67 |
| 11 | 1.31 | 3.93 | 156.44 | 48.74 |
| 12 | 1.31 | 3.14 | 142.58 | 51.54 |
| 13 | 0.79 | 3.14 | 15.30 | 61.08 |
| 14 | 0.00 | 3.14 | -125.87 | 55.51 |
| 15 | 0.00 | 2.36 | -122.96 | 43.33 |
| 16 | 0.79 | 2.36 | 19.22 | 43.93 |
| 17 | 1.31 | 2.36 | 139.83 | 44.98 |
| 18 | 1.31 | 1.83 | 143.35 | 38.07 |
| 19 | 2.36 | 1.83 | 26.36 | 31.11 |
| 20 | 3.14 | 1.83 | -126.43 | 30.20 |
| 21 | 3.14 | 0.79 | -124.84 | 46.20 |
| 22 | 2.36 | 0.79 | 30.06 | 41.52 |
| 23 | 1.31 | 0.79 | 150.05 | 47.92 |
| 24 | 1.31 | 0.00 | 150.01 | 51.39 |

Table S4. Parameters on path $\Gamma_4$.

| Points No. | $\phi_x$ (rad) | $\phi_y$ (rad) | X (rad/s) | Y (rad/s) |
|---|---|---|---|---|
| 1 | 6.28 | 5.50 | -122.08 | 47.44 |
| 2 | 4.97 | 5.50 | 150.60 | 39.35 |
| 3 | 5.50 | 4.97 | 29.11 | 38.34 |
| 4 | 6.28 | 4.97 | -122.94 | 34.79 |
| 5 | 6.28 | 3.93 | -115.76 | 46.90 |
| 6 | 5.50 | 3.93 | 28.85 | 47.33 |
| 7 | 4.97 | 3.93 | 154.73 | 44.23 |
| 8 | 4.97 | 3.14 | 146.22 | 60.48 |
| 9 | 5.50 | 3.14 | 29.25 | 50.58 |
| 10 | 6.28 | 3.14 | -125.87 | 55.51 |
| 11 | 6.28 | 2.36 | -122.96 | 43.33 |
| 12 | 5.50 | 2.36 | 28.16 | 38.99 |
| 13 | 4.97 | 2.36 | 147.39 | 45.34 |
| 14 | 4.97 | 1.83 | 156.05 | 34.52 |
| 15 | 5.50 | 1.83 | 27.57 | 31.17 |
| 16 | 6.28 | 1.83 | -127.55 | 30.02 |
| 17 | 6.28 | 0.79 | -126.94 | 51.91 |
| 18 | 5.50 | 0.79 | 23.56 | 40.13 |
| 19 | 4.97 | 0.79 | 138.11 | 42.40 |
| 20 | 4.97 | 0.00 | 151.35 | 57.68 |
| 21 | 3.93 | 0.00 | 25.47 | 53.04 |
| 22 | 3.14 | 0.00 | -130.79 | 52.12 |
| 23 | 3.14 | 0.79 | -124.84 | 46.20 |
| 24 | 2.36 | 0.79 | 30.06 | 41.52 |
| 25 | 1.31 | 0.79 | 150.05 | 47.92 |
| 26 | 1.31 | 0.00 | 150.01 | 51.39 |
| 27 | 0.79 | 0.00 | 27.31 | 50.89 |

**Table S5.** Parameters on paths $\Gamma_5$ and $\Gamma_6$.

| Points No. | $\phi_x$ (rad) | $\phi_y$ (rad) | Path $\Gamma_5$ | | Path $\Gamma_6$ | |
|---|---|---|---|---|---|---|
| | | | X (rad/s) | Y (rad/s) | X (rad/s) | Y (rad/s) |
| 1 | 1.83 | 0.00 | 143.02 | 56.98 | 141.14 | 59.17 |
| 2 | 1.83 | 0.79 | 144.00 | 44.05 | 137.64 | 38.78 |
| 3 | 1.83 | 1.31 | 153.75 | 30.70 | 142.58 | 31.28 |
| 4 | 1.83 | 2.36 | 154.74 | 44.38 | 149.74 | 39.27 |
| 5 | 2.62 | 2.36 | -46.87 | 43.19 | -49.44 | 42.07 |
| 6 | 3.14 | 2.36 | -125.27 | 42.44 | -121.55 | 41.80 |
| 7 | 2.62 | 3.14 | -49.93 | 59.47 | -49.71 | 53.72 |
| 8 | 1.83 | 3.14 | 143.78 | 58.06 | 145.32 | 59.98 |
| 9 | 1.83 | 3.93 | 141.45 | 38.69 | 154.26 | 44.87 |
| 10 | 1.83 | 4.45 | 139.82 | 30.95 | 143.33 | 29.62 |
| 11 | 1.83 | 4.97 | 151.06 | 28.47 | 158.84 | 35.61 |
| 12 | 1.83 | 5.50 | 149.49 | 45.95 | 158.26 | 40.92 |
| 13 | 2.62 | 5.50 | -43.03 | 39.76 | -58.38 | 39.07 |
| 14 | 3.14 | 5.50 | -123.33 | 41.71 | -126.76 | 42.28 |
| 15 | 3.14 | 6.28 | -122.08 | 55.23 | -132.75 | 52.44 |
| 16 | 2.62 | 6.28 | -45.26 | 50.66 | -58.29 | 51.23 |
| 17 | 1.83 | 6.28 | 143.02 | 56.98 | 141.14 | 59.17 |

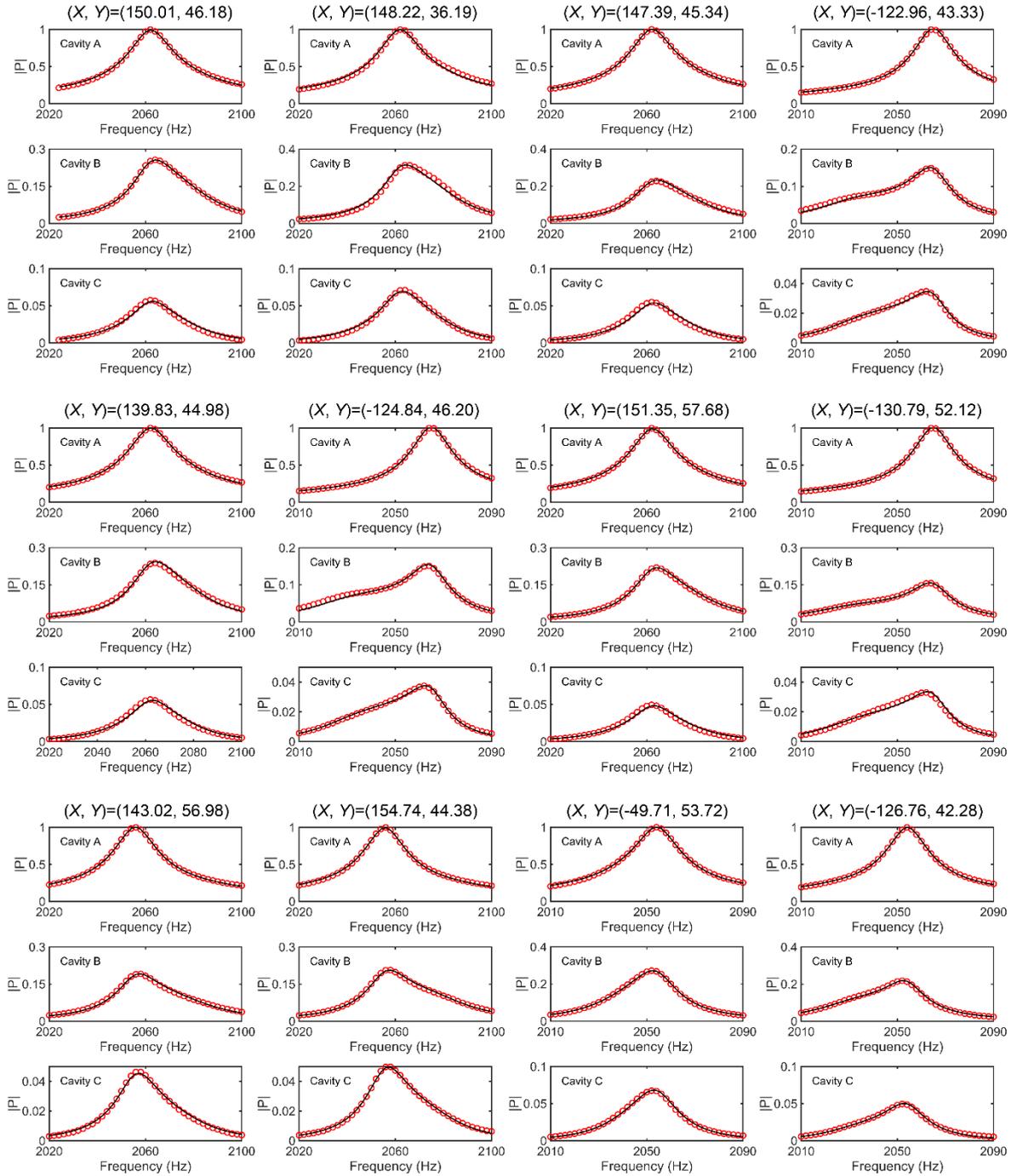

**Fig. S4.** Selected data on paths $\Gamma_1$ to $\Gamma_6$. Good agreement is seen between the experimental results (red open markers) and the fitted results (black solid curves).